\documentclass[a4paper,11pt]{article}

\usepackage{jinstpub} 
\RequirePackage{xspace}
\def\epem {\ensuremath{e^+e^-}\xspace}

\def\gg {\ensuremath{\gamma \gamma}\xspace}
\def\ge {\ensuremath{\gamma e}\xspace}

\def\t-t {\ensuremath{{\rm   tt}}\xspace}

\newcommand{\mev}{\ensuremath{\mathrm{\,Me\kern -0.1em V}}\xspace}
\newcommand{\mevcc}{\ensuremath{{\mathrm{\,Me\kern -0.1em V\!/}c^2}}\xspace}
\newcommand{\gev}{\ensuremath{\mathrm{\,Ge\kern -0.1em V}}\xspace}
\newcommand{\gevcc}{\ensuremath{{\mathrm{\,Ge\kern -0.1em V\!/}c^2}}\xspace}
\newcommand{\tev}{\ensuremath{\mathrm{\,Te\kern -0.1em V}}\xspace}
\newcommand{\tevcc}{\ensuremath{{\mathrm{\,Te\kern -0.1em V\!/}c^2}}\xspace}
\newcommand{\ev}{\ensuremath{\mathrm{\,e\kern -0.1em V}}\xspace}

\def\cm   {\ensuremath{{\rm \,cm}}\xspace}

\def\mum  {\ensuremath{{\,\mu\rm m}}\xspace}

\newcommand{\EMEM}{\ensuremath{e^-e^-}\xspace}
\newcommand{\GG}{\ensuremath{\gamma\gamma}\xspace}
\newcommand{\GE}{\ensuremath{\gamma e}\xspace}
\newcommand{\LGG}{\ensuremath{L_{\gamma\gamma}}}

\newcommand{\mrad}{\ensuremath{\rm \,mrad}\xspace}

\newcommand{\be}{\begin{equation}}
\newcommand{\ee}{\end{equation}}
\newcommand{\bc}{\begin{center}}
\newcommand{\ec}{\end{center}}
\newcommand{\bi}{\begin{itemize}}
\newcommand{\ei}{\end{itemize}}
\newcommand{\ben}{\begin{enumerate}}
\newcommand{\een}{\end{enumerate}}

\title{\boldmath Optimization of the beam crossing angle at the ILC for \epem and \gg collisions}

\author[a,b]{V.~I.~Telnov}

\affiliation[a]{Budker Institute of Nuclear Physics, Novosibirsk, Russia}
\affiliation[b]{Novosibirsk State University, Novosibirsk, Russia}

\emailAdd{telnov@inp.nsk.su}

\abstract{At this time, the design of the International Linear Collider (ILC) is optimized for \epem collisions; the photon collider (\gg and \ge) is considered as an option. Unexpected discoveries, such as the diphoton excess $\digamma(750)$ seen at the LHC, could strongly motivate the construction of a photon collider. In order to enable the \gg collision option, the ILC design should be compatible with it from the very beginning. In this paper, we discuss the problem of the beam crossing angle. In the ILC technical design~\cite{ILC}, this angle is 14 \mrad, which is just enough to provide enough space for the final quadrupoles and outgoing beams. For \gg collisions, the crossing angle must be larger because the low-energy electrons that result from multiple Compton scattering get large disruption angles in collisions with the opposing electron beam and some deflection in the solenoidal detector field. For a $2E_0=500 \gev$ collider, the required crossing angle is about $25 \mrad$. In this paper, we consider the factors that determine the crossing angle as well as its minimum permissible value that does not yet cause a considerable reduction of the \gg luminosity. It is shown that the best solution is to increase the laser wavelength from the current $1 \mum$ (which is optimal for $2E_0=500 \gev$) to $2 \mum$ as this makes possible achieving high \gg luminosities at a crossing angle of $20 \mrad$, which is also quite comfortable for \epem collisions, does not cause any degradation of the \epem luminosity and opens the possibility for a more energetic future collider in the same tunnel (e.g., CLIC). Moreover, the $2 \mum$ wavelength is optimal for a $2E_0 = 1 \tev$ collider, e.g., a possible ILC energy upgrade. Please consider this paper an appeal to increase the ILC crossing angle from 14 to $20 \mrad$.}


\arxivnumber{1801.10471} 

\begin{document}
\maketitle
\flushbottom
\section{Introduction}\label{s1}
   Linear \epem colliders are generally considered to be the best tool for detailed study of new physics at energies $2E_0 \approx 0.2$--$3 \tev$. They have been actively developed since 1980s. One of the projects, the ILC~\cite{ILC}, for energy $2E_0=250$--$500 \gev$ (and the potential to be upgraded to $1 \tev$), is ready for construction in Japan and awaits approval. Development of another project, CLIC~\cite{CLIC}, for energy up to $3 \tev$, is also near completion. The future of both projects depends on the energy scale of new physics. Precision study of the Higgs boson and the top quark is a rather good motivation for the ILC(500), while for ILC(1000) and CLIC additional motivation is needed; LHC experiments may help guide the decision.

   The photon linear collider (PLC) based on conversion of linear-collider electrons to high-energy photons using Compton scattering of laser photons has been discussed and actively developed since the early 1980s~\cite{GKST81,GKST83}. A PLC would be a very natural and relatively inexpensive addition to a high-energy \epem linear collider. The PLC would enable the study of new physics in two additional types of collisions, \gg and \ge, at energies and luminosities close to those in \epem collisions.  Nearly every aspect of the comprehensive description of the PLC in the TESLA TDR~\cite{TESLATDR} is also valid for the PLC at the ILC, which we discuss below.
\begin{figure}[!htb]
 \vspace{-0.cm}
\begin{center}
\hspace{0.0cm} \includegraphics[width=10.cm,height=9.cm]{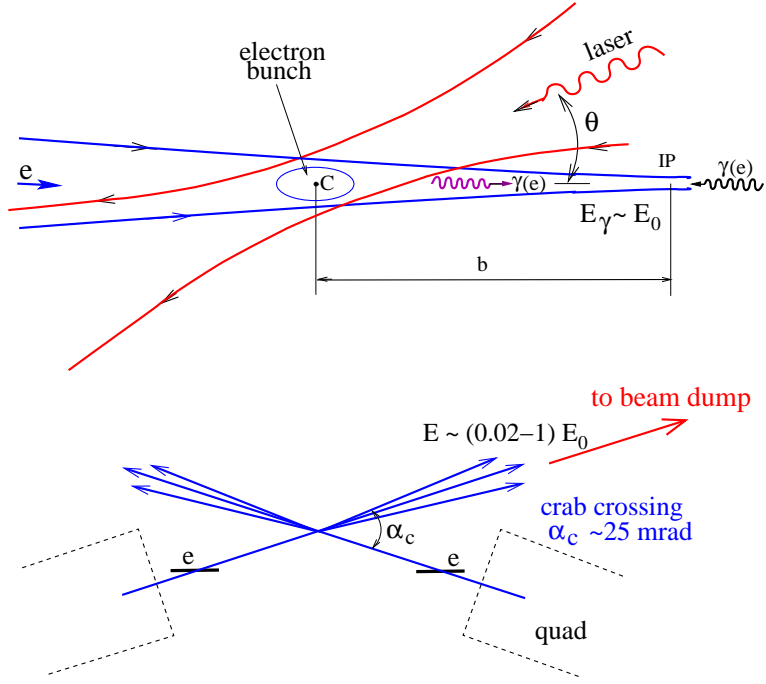}
\caption{\emph{(top)} The general scheme of a \gg, \ge photon collider. \emph{(below)} A crab-crossing collision scheme for the removal of disrupted beams from the detector to the beam dump. }
\end{center}
\label{scheme}
\vspace{-0.5cm}
\end{figure}

   The general PLC scheme is shown in Fig.~\ref{scheme} (top). Two electron beams, each of energy $E_0$, after passing the final focus system travel
towards the interaction point (IP). At a distance $b \sim \gamma \sigma_y$, or about 1--3 mm, from the IP, they collide with a focused laser beam.  After Compton scattering, the photons have an energy close to that of the initial electrons and follow their direction to the IP (with a small additional angular spread of the order of $1/\gamma$, where $\gamma =E_0/mc^2$), where they collide with an opposing beam of high-energy photons or electrons.  Using a picosecond laser with a flash energy of 5--10 joules, one can ``convert'' almost all electrons to high-energy photons. The photon spot size at the IP will be almost equal to that of the electrons, and therefore the total luminosity in \gg,
\ge collisions will be similar to the ``geometric'' luminosity of the underlying electron beams.

In the conversion region, a laser photon of energy $\omega_0$
collides with a high-energy electron of energy $E_0$ at a small
collision angle (almost head-on).  The energy of the
scattered photon $\omega$ depends on the photon scattering angle
$\vartheta$ with respect to the initial direction of the electron as
follows~\cite{GKST83}: \vspace*{-1.3mm}
\begin{equation}
\omega = \frac{\omega_\mathrm{m}}{1+(\vartheta/\vartheta_0)^2},\;\;\;\;
\omega_\mathrm{m}=\frac{x}{x+1+\xi^2}E_0, \;\;\;\;
\vartheta_0= \frac{mc^2}{E_0} \sqrt{x+1},
\label{kin}
\end{equation} \vspace*{-0.8mm}
\noindent where  \vspace*{-0.9mm}
\begin{equation}
x \simeq \frac{4E \omega_0 }{m^2c^4}
 \simeq 15.3\left[\frac{E_0}{\tev}\right]
\left[\frac{\omega_0}{\ev}\right] =
 19\left[\frac{E_0}{\tev}\right]
\left[\frac{\mu \mbox{m}}{\lambda}\right], \label{e3:x}
\end{equation}
\noindent where $\omega_\mathrm{m}$ is the maximum energy of scattered photons and $\xi^2$ is the parameter characterizing nonlinear effects in Compton scattering~\cite{TESLATDR}. Tighter focusing of the laser beam reduces the required flash energy but leads to a decreased maximum energy of scattered photons.
In order to keep the energy shift below 5\%, $\xi^2$ should be in the 0.15--0.3 range for typical values $x=2$--5.
The maximum value of $x$ is about 4.8 ($\lambda \approx 4.2 E_0 \;[\tev]\; \mum$) due to \epem  pair creation in collisions of high-energy and laser photons in the conversion region~\cite{GKST83,TEL90,TEL95}, see also section~\ref{2mum}. Therefore, the maximum collision energy
is about 80\% for \GG and 90\% for \ge collisions.
For example, if $E_0 = 250 \gev$, $\omega_0 = 1.17 \ev$ ($\lambda=1.06 \mum$, for the most powerful solid-state lasers), then $x=4.5$
and $\omega_\mathrm{m}/E_0 = 0.82$.  Formulae for the Compton cross section can
be found elsewhere~\cite{GKST83,GKST84,TESLATDR}.

After crossing the conversion region, the electrons have a very broad energy spectrum, $E=($0.02--1)\,$E_0$, and large disruption angles due to deflection of low-energy electrons in the fields of the opposing beam and the detector solenoid. The removal of such a beam from the detector is therefore far from trivial.

The ``crab crossing'' scheme of beam collisions solves the problem of beam removal at photon colliders~\cite{TEL90, TEL95}, Fig.~\ref{scheme} (bottom).  In the crab-crossing scheme~\cite{Palmer}, the beams are collided at a crossing angle $\alpha_\mathrm{c}$.  In order to preserve the
luminosity, the beams are tilted by a special RF cavity by the angle $\alpha_\mathrm{c}/2$.  If the crossing angle is larger than the disruption angles, the beams just travel straight outside the quadrupole magnets.

\begin{figure}[!htb]
\begin{minipage}{0.5\linewidth}
 \vspace{-0.8cm}
  \hspace{-0.5cm}\includegraphics[width=8.5cm,height=8.cm]{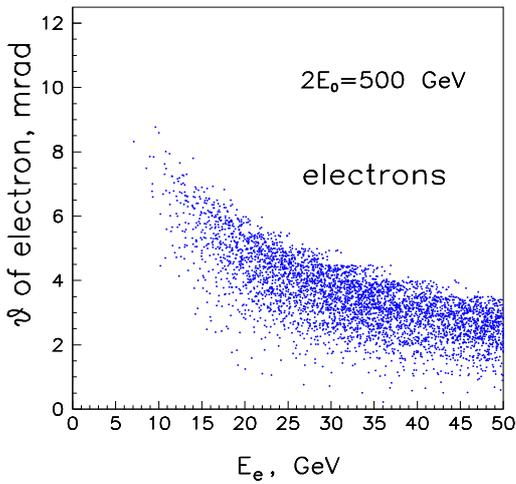}
\end{minipage} \hspace{0.5cm}
\begin{minipage}{0.43\linewidth} \hspace{0mm} \vspace{-1.8cm}
\caption{Angles of disrupted electrons after Compton
    scattering and interaction with the opposing electron beam; $N=2\times
    10^{10}$, $\sigma_z=0.3$ mm.}  \hspace{0.5cm}
\label{e-angle}
\end{minipage}
\vspace{-1.cm}
\end{figure}

The dependence of the disruption angle on the electron energy, obtained by simulation, is shown in Fig.~\ref{e-angle}~\cite{PHOTON2005b}. After passing the conversion and collision points, the electrons have
energy ranging from about $5 \gev$ to $E_0$, and the horizontal disruption
angle can be as large as $10 \mrad$, see Fig.~\ref{e-angle}. Above this angle, the total energy of particles is less than that in
the secondary irremovable \epem background. The disruption angle for low-energy particles is proportional to $\sqrt{N/\sigma_z  E}$~\cite{TEL90,TEL95} and depends very weakly on the transverse beam size, see section~\ref{s-def}.

\begin{figure}[!htb]
\vspace{-1cm}
\hspace{-0.0cm} \includegraphics[width=7.7cm]{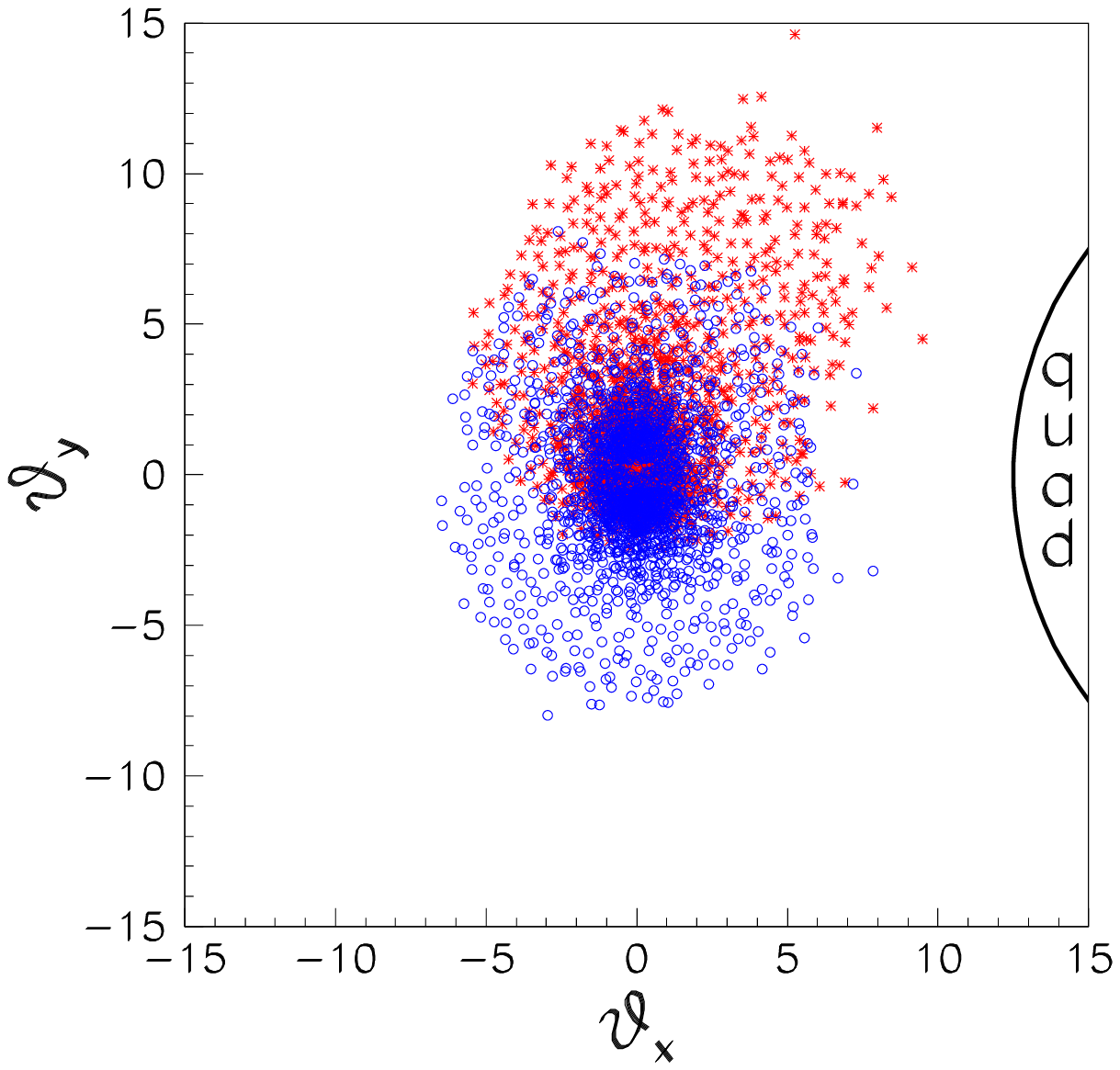}
\includegraphics[width=7.7cm]{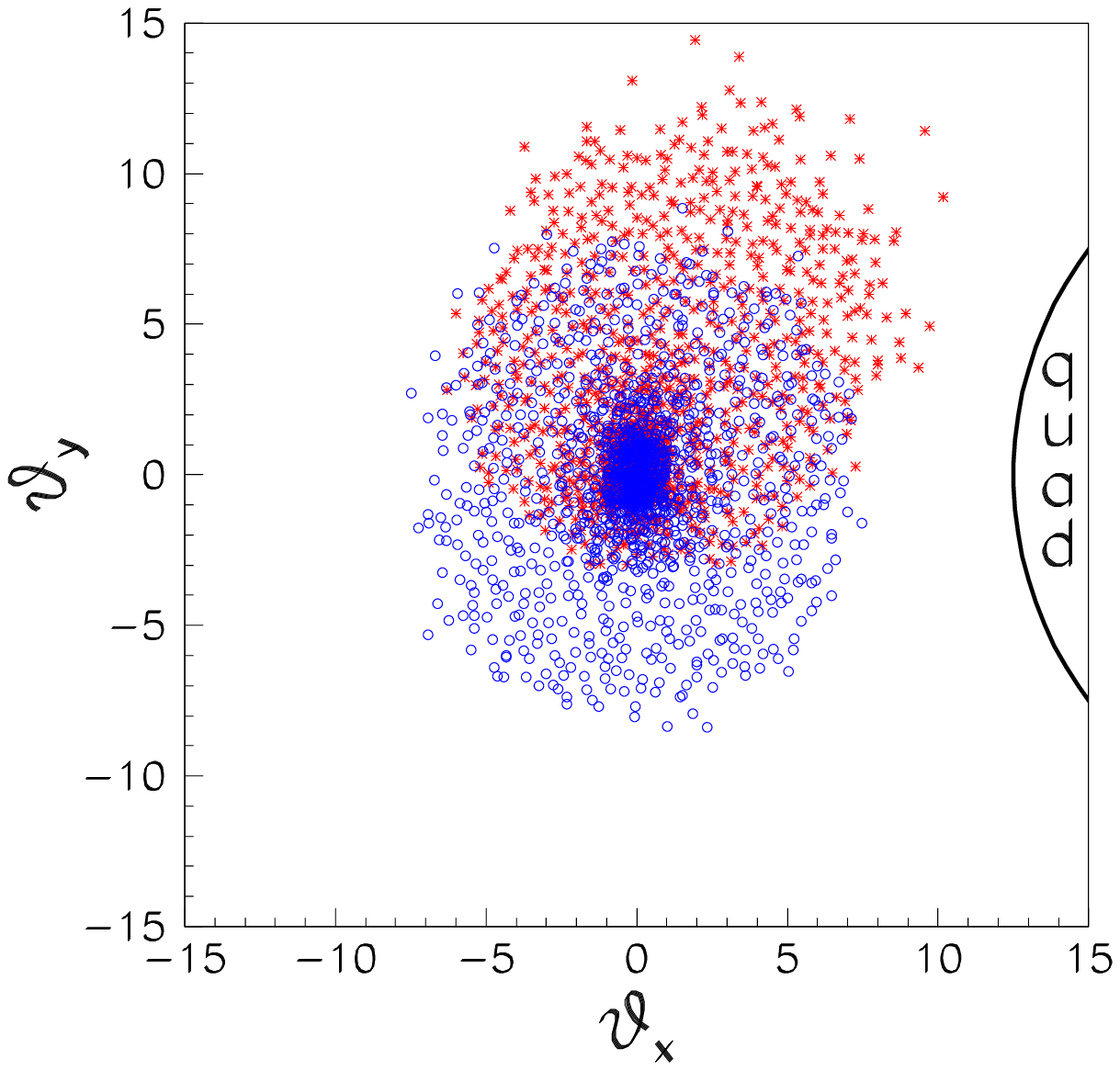} \vspace{-1.4cm}
\caption{The cross section of the disrupted beam at the distance of $4 \,\mathrm{m}$ from the IP, at the place where they
  pass the first quad.  The shift of the outgoing beam due to the detector field is seen. Blue (square) points: only beam-beam deflection; red (star) points: the detector field of $4 \,\mathrm{T}$ is added.  Left figure: $2E_0=200 \gev$, right: $2E_0=500 \gev$. }
\label{thxthy} \vspace{-0.0cm}
\end{figure}

  Due to the crossing angle, the detector field gives an additional deflection angle to the disrupted beam, see Fig.~\ref{thxthy}.

The required crossing angle is determined by the disruption angle, the outer radius of the final quadrupole magnet (about $5 \cm$~\cite{PHOTON2005b}), and the distance between the first quadrupole and the IP (about $4 \,\mathrm{m}$), which gives
$\alpha_\mathrm{c} \approx 12 \mrad + 5/400 \approx 25 \mrad$.

\begin{figure}[!htb]
\begin{minipage}{0.5\linewidth}
 \vspace{-0.cm}
\hspace{10mm}\includegraphics[width=6.cm,bb=0 0 482 475, clip]{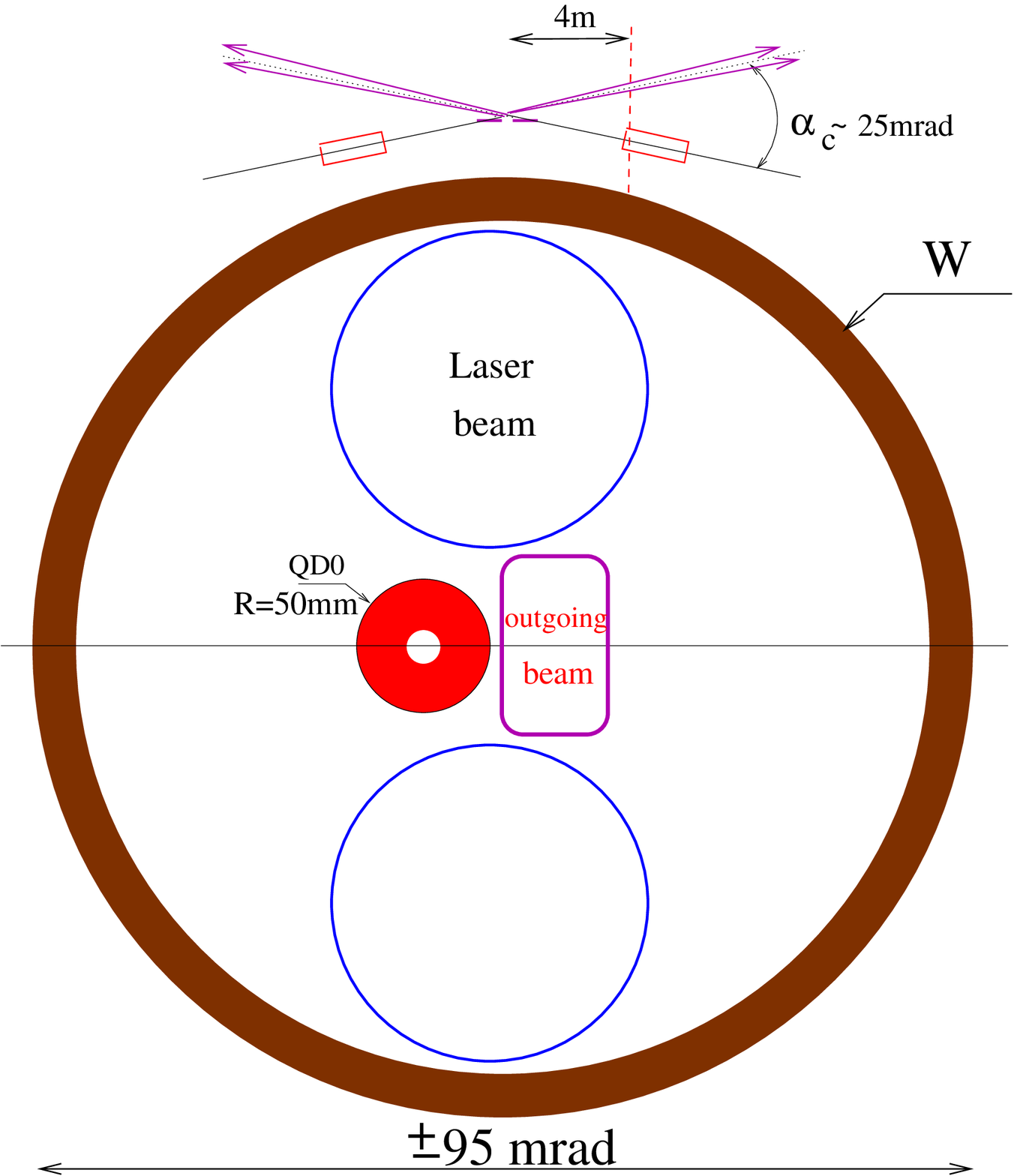}
\end{minipage} \hspace{-0.cm}
\begin{minipage}{0.47\linewidth}
\caption{Layout of the quadrupole magnet and electron and laser beams at the
  distance of $4 \,\mathrm{m}$ from the interaction point.}
\label{beams-quad}\end{minipage}
\end{figure}

The layout of the final quad, the outgoing electron beam and the
laser beams at the distance $4 \,\mathrm{m}$ from the IP is shown in
Fig.~\ref{beams-quad}.

The current ILC design~\cite{ILC} is optimized for \epem collisions: only one IP is planned, with two detectors in a push-pull configuration and a crossing angle of $14 \mrad$, which is a minimal angle that provides just enough space for the final quadrupoles and the outgoing \epem beam lines. At the photon collider, the beam-crossing angle must be no less than $25 \mrad$. Therefore, at present, the ILC design is incompatible with the photon collider. This issue was known from the early days of ILC R\&D; however, attempts to correct it and make the ILC design compatible with the photon collider have been sidelined by overriding cost concerns: in order to minimize the cost of the ILC project and increase the chances of its approval, only \epem collisions were included in the baseline and technical ILC designs.

Another reason this issue is still pending is the absence, as of yet, of a compelling physics case for the \gg collider. In particular, the Higgs boson can be studied much better in \epem collisions; the photon collider is preferable only for the measurement of the Higgs boson's gamma-gamma width~\cite{Tel2016}. The photon collider can be helpful in many ``new physics'' scenarios---but, unfortunately, no new physics has yet been seen at the LHC.

   In 2015, two detectors at the LHC saw evidence for a diphoton peak at $W \approx 750 \gevcc$, which caused a great deal of excitement in the HEP community, resulted in more than 500 papers, but in the end was proven to be a statistical fluctuation. The ILC--CLIC director Lyn Evans wrote in LC Newsline: ``On the scientific side, there was much discussion of the possible sighting of a new resonance at $750 \gev$ at the LHC and its implications for the ILC. If this resonance is confirmed in the coming months, it is recommended that the possible option of running the ILC as a gamma-gamma collider at $1 \tev$ as well as an \epem collider be strongly pursued. This would require a minor modification of the ILC layout.''

   The critical point here is that the ILC design must be modified (the crossing angle increased, a special beam dump~\cite{S-T} added) before the construction starts; doing it at a later stage would be practically impossible.

     It is said that God likes to speak to us indirectly, by dropping hints. If so, this short-lived diphoton bump may have been a gentle reminder to us all that it is high time we correct a major omission in the ILC design and make it compatible with the photon collider.

     So, a beam crossing angle of $14 \mrad$ is too small for \gg collisions at the ILC, while an angle of $25 \mrad$ is optimal. Now, let us consider the reasons for keeping the crossing angle as small as possible from the point of view of \epem collisions. The are two primary considerations:
   \begin{enumerate}
    \item a smaller crossing angle improves detector hermeticity, which may be important for certain SUSY searches that require suppression of two-photon backgrounds by tagging the scattered electrons. Here, as has been shown by simulations performed more than a decade ago, an increase of the crossing angle leads only to a relatively small decrease in detection efficiency;
    \item too large a crossing angle may lead to broadening of the vertical spot size at the IP due to synchrotron radiation in the detector solenoid. Simulations performed in 2005~\cite{telnov-2005-cross} for three ILC detector proposals, LDC, SID and GLD, are shown in table~1. This effect is rather small for $\alpha_\mathrm{c} < 25 \mrad$.
 {
 \vspace{-.1cm}
\begin{table}[ht]
\bc
\caption{ Results on  $L(\alpha_\mathrm{c})/L(0)$ for \epem collisions.}  \setlength{\tabcolsep}{3.mm}
\vspace{2mm}
\begin{tabular}{lllllll}
$\alpha_\mathrm{c}$(\mrad) & 0 & 20 & 25 & 30 & 35 & 40 \\ \hline
LDC & 1. &0.997 &0.995 &0.99 &0.985 & 0.973 \\
SID & 1. &0.997 &0.993 &0.985 &0.97 & 0.93 \\
GLD & 1. &0.995 &0.99 &0.98 &0.96  & 0.935 \\
\end{tabular}
\ec
\vspace{-.7cm}
\end{table}
}
   \end{enumerate}
   The goal of the present paper is to study the effect of reducing the crossing angle to a value below $25 \mrad$ on the photon collider luminosity and to suggest a compromise value of the crossing angle that provides for acceptably high luminosities in both \epem and \gg collisions.

\section{Factors determining the disruption angle}

\paragraph{Minimum energy.}
   The maximum energy $\omega_\mathrm{m}$ of scattered photons in Compton scattering is given by Eq.\ref{kin}.
The minimum energy of the electron after one scattering is $E_{{\rm min},1}=E_0/(x+1)$. Taking into account the fact that the value of $x$ decreases after each scattering, we obtain the minimum energy after $n$ scatterings~\cite{GKST83}
\be
E_{\rm min,n}=\frac{E_0}{nx+1} \approx \frac{E_0}{nx},
\label{emin1}
\ee
here $x$ is the initial value of $x$, optimally $x\sim 4.8$, and $n \sim$ 5--10.

\begin{figure}[!htb]
\begin{minipage}{0.5\linewidth}
\includegraphics[width=8cm]{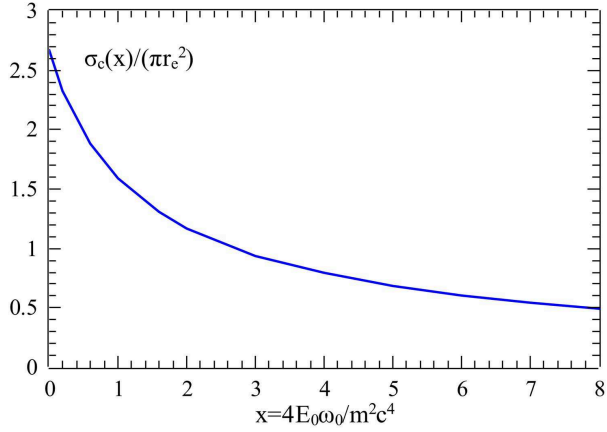}
\end{minipage} \hspace{0.5cm}
\begin{minipage}{0.43\linewidth} \hspace{0mm} \vspace{-1.0cm}
\caption{Dependence of the Compton cross section on the parameter $x$}  \hspace{0.5cm}
\label{cross-sec}
\end{minipage}
\vspace{-0.5cm}
\end{figure}

The cross section of Compton scattering depends on $x$ as shown in Fig.~\ref{cross-sec}. It is equal to the Thompson cross section $\sigma_\mathrm{T}=(8/3)\pi r_e^2$ at $x=0$ and decreases to $0.25 \sigma_\mathrm{T}$ at $x=5$.
   In the conversion region, the cross section almost reaches its maximum value after the first backward Compton scattering. The probability of Compton scattering of an electron with the initial energy $p \sim \sigma_\mathrm{c} n_{\gamma}l$, where $n_{\gamma}$ is the photon density and $l$ is the thickness of the laser target; usually, we assume $p \sim 1$. The average number of Compton scatterings by the same electron $n\sim p \sigma_\mathrm{T}/\sigma_\mathrm{c}(x)$. The maximum number of electron scatterings (which is important for the backgrounds) is proportional to this average $n$. Substituting $n$ to Eq.~\ref{emin1}, we find the minimum energy of electrons after multiple Compton scattering
  \be
  E_{\rm min} \sim \frac{E_0}{nx} \propto \frac{m^2 c^4 \sigma_\mathrm{c}(x)}{\omega_0 p \sigma_\mathrm{T}}.
  \label{emin2}
  \ee

\paragraph{Deflection angle.} \label{s-def}
    Consider an electron that has energy $E$ after passing the conversion region and collides with the opposing electron beam. At linear colliders, beams are flat. The radius of curvature $R_0 \propto E\sigma_x \sigma_z/e^2N$ (numerical coefficients are omitted) at $r < \sigma_x$ and $R \approx R_0 r/ \sigma_x$ at $r > \sigma_x$. When the electron gets a vertical displacement equal to the horizontal beam size ($\sigma_x=R_0\theta_0^2/2$) and leaves the volume with a strong field, its deflection angle
    \be
     \theta_0 \sim \sqrt{\frac{2e^2N}{\sigma_z E}}.
    \label{theta-0}
    \ee
Electron's motion outside the beam is described by the equations
\be \mathrm{d}\theta = \frac{\mathrm{d}z}{R}= \frac{\mathrm{d}z\, \sigma_x}{R_0 r}, \;\;\;\;\;\;\; \mathrm{d}r= \theta\, \mathrm{d}z. \ee
Excluding $\theta$, we obtain
\be
\ddot{r}=\frac{\sigma_x}{R_0 r},   \;\;\; {\rm or} \;\;\; \dot{r}^2= \frac{2\sigma_x}{R_0} \ln{r} + {\rm const}.
\ee
With the initial conditions  $\theta=\theta_0$ at $r=\sigma_x$,
\be
\dot{r}^2 \equiv \theta^2 =\frac{2\sigma_x}{R_0} \ln{\frac{r}{\sigma_x}} + \theta_0^2= \theta_0^2\left(1+\ln{\frac{r}{\sigma_x}}\right).
\ee
This equation has no analytical solution; however, it can be solved approximately.
Due to weak logarithmic dependence on $r_{\rm max}$, one can take, as a first approximation, $r_{\rm max}=\sigma_z\theta_0$; then,
the total deflection angle
 \be
\theta_{\rm max}^2 \sim \theta_0^2\left(1+\ln{\frac{\theta_0 \sigma_z}{\sigma_x}}\right).
\ee
Thus, with logarithmic accuracy, the deflection angle is simply proportional to $\theta_0$ as given by Eq.~\ref{theta-0}
and does not depend on the transverse beam sizes.

 After substitution of $E$ from Eq.~\ref{emin2} to Eq.~\ref{theta-0} we finally obtain the value of the disruption angle
 \be
 \theta_\mathrm{d} \propto \sqrt{\frac{Np\omega_0}{\sigma_z \sigma_c(x)}} \propto \sqrt{\frac{Np}{\sigma_z \sigma_c(x) \lambda}}.
 \ee
 As the Compton cross section $\sigma_\mathrm{c}$ decreases with the increase of $x$, we can conclude that at the fixed laser wavelength and conversion probability $p$, the disruption angle reaches its maximum value at the highest collider energy.

\section{Ways to reduce the crossing angle}
\paragraph{\boldmath \gg luminosity.}
The \gg luminosity is proportional to the geometric \EMEM luminosity:
\be
\LGG \propto k^2 \frac{N^2f}{\sigma_x\sigma_y},
\ee
where $k \approx 1-e^{-p}$, $\sigma_y \propto \sqrt{\beta_y} \approx \sqrt{\sigma_z}$, and the product $Nf$ is determined by the RF power of the collider and is fixed.

\begin{figure}[!htb]
\centering
\hspace{-1.0cm} \includegraphics[width=8.0cm, bb=53 70 431 404, clip]{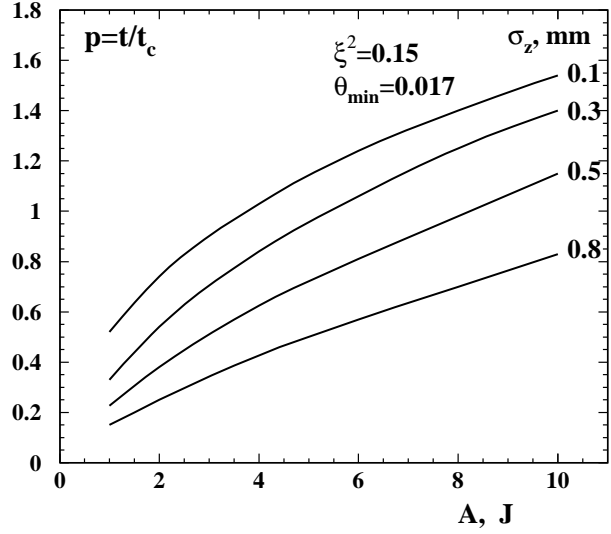}\hspace{-1cm}
\vspace{-0.1cm}
\caption{Dependence of the conversion probability on the electron bunch length and laser flash energy. }
\label{apsz} \vspace{0.cm}
\end{figure}

Our goal is to find a way to decrease the disruption angle with a minimal decrease of the \gg luminosity. We have
 \be
 \LGG \propto \frac{N (1-e^{-p})^2}{\sqrt{\sigma_z}} \sim \frac{N p^{1.15}}{\sqrt{\sigma_z}}, \;\;{\rm at} \;\; p \approx 1;
 \ee
\be
\theta_\mathrm{d} \propto \sqrt{\frac{Np}{\sigma_z \lambda}}.
\label{d-angle}
\ee
Up to now, the recommended crossing angle for the photon collider has been $\alpha_\mathrm{c}=25 \mrad$, with roughly one half of it determined by the size of the quadrupole magnets and another half by the disruption angle. In order to reduce $\alpha_\mathrm{c}$ from 25 to $20 \mrad$, we have to reduce $\theta_\mathrm{d}$ by 5 \mrad, or by a factor of $12.5/7.5=1.67$.
\paragraph{Number of particles.}
Obviously, it makes no sense to decrease $N$ because in this case $\LGG \propto \theta_\mathrm{d}^2$; besides, it would be difficult to increase the collision rate to keep the product $Nf$ constant.
\paragraph{Conversion probability.}
Reducing the collision probability in the conversion region, $p$, also leads to a considerable loss of the \gg luminosity: $\LGG \propto \theta_\mathrm{d}^{2.3}$.
\paragraph{Bunch length.}
 It looks more reasonable to increase the electron bunch length, as in this case $\LGG \propto \theta_\mathrm{d}$. However, to keep the conversion probability constant one would need to increase the laser flash energy $A$. Its dependence on the bunch length is shown in Fig.~\ref{apsz}. In order to decrease the disruption angle by the desired factor of 1.67, the bunch length would have to be increased by a factor of $(1.67)^2=2.8$, which requires a factor 3 greater laser flash energy (for $p= \rm const$). Laser power is a leading challenge for the photon collider, and thus such a big increase is excluded.

\begin{figure}[!htb]
\vspace{0.0cm}
\hspace{-0.0cm} \includegraphics[width=7.cm]{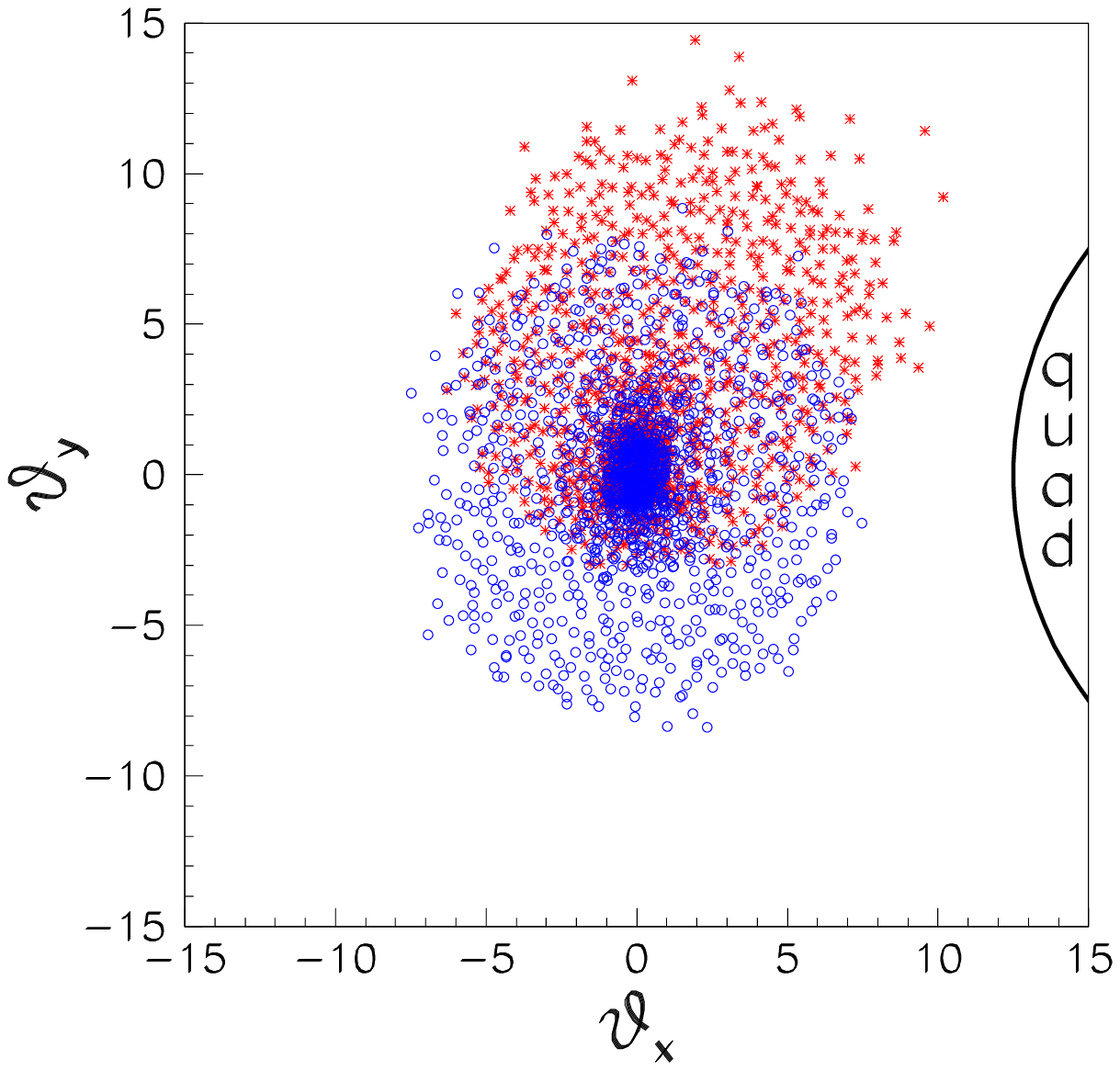}
\includegraphics[width=7.cm]{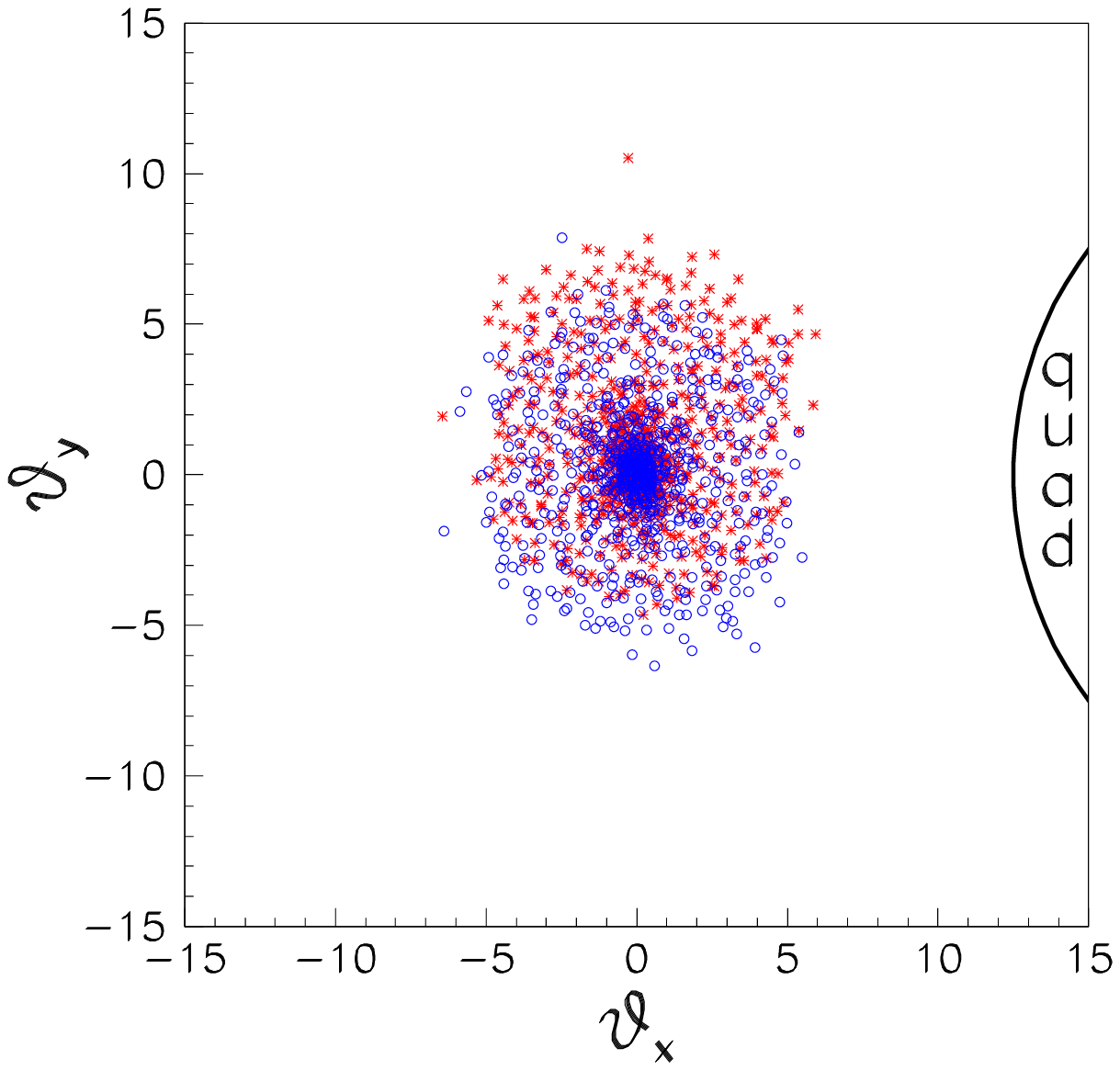}
\vspace{0.1cm}
\caption{The cross section of the disrupted beams at the distance of $4 \,\mathrm{m}$ from the IP, where they
  pass the first quad (similar to Fig.~\ref{thxthy}). Blue (square) points: only beam-beam deflection, red (stars)  points: same with a detector field of $4 \,\mathrm{T}$.  Left figure: $2E_0=500$ GeV, $\lambda=1$ \mum; right: $2E_0=1000$ GeV, $\lambda=2$ \mum.}
\label{thxthy2}
\end{figure}

\paragraph{Laser wavelength.}
 Last but not least, one can increase the laser wavelength. According to Eq.~\ref{d-angle}, the disruption angle drops as $1/\sqrt{\lambda}$ without any change in the \gg luminosity. So, if we take $\lambda=2 \mum$ instead of the current $\lambda=1 \mum$, the disruption angle would be a factor 1.41 smaller. Moreover, as shown in Fig.~\ref{thxthy}, the total disruption angle consists of two parts: the disruption angle due to the collision with the opposing electron beam plus an additional deflection in the solenoidal field of the detector, which is proportional to $1/E$. According to Eq.~\ref{emin2}, the minimum electron energy is proportional to the laser wavelength; therefore the deflection angle due to the detector field would be reduced by a factor of 2. These effects are clearly seen in Fig.~\ref{thxthy2}, where the right figure corresponds to $2E_0=1000 \gev$, $\lambda=2 \mum$ and the left one to $2E_0=500 \gev$, $\lambda=1 \mum$. One can see a considerable shrinking of the transverse beam sizes at $\lambda = 2 \mum$, sufficient to reduce the crossing angle to $20 \mrad$.

 Please note that the energy in the figure on the right is taken to be a factor of two greater than in the figure on the left. The reason for this is the fact that a 2 \mum laser would allow working with electron energies twice as high without creating \epem pairs in the conversion region (threshold $x=4.8$). The parameter $x=4.75$ for both figures.  As shown in section~\ref{s-def}, the maximum value of  the disruption angle is achieved at the maximum collider energy, so these two cases are the worst cases for these two wavelengths.
\vspace{-0.1cm}

 \section{Advantages of transitioning from {\boldmath $\lambda=1 \mum$  to $\lambda=2 \mum$}.}
\label{2mum}
\begin{figure}[!htb]
\centering
\hspace{-1.0cm} \includegraphics[width=10.0cm]{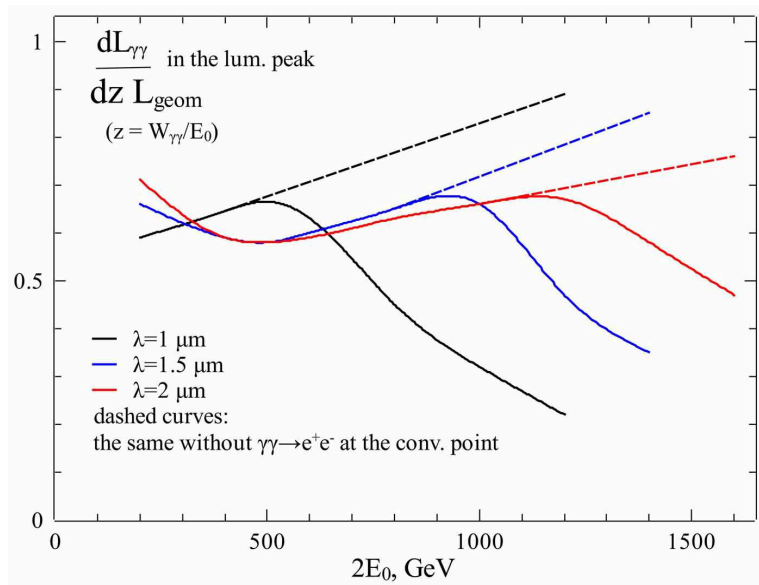}\hspace{-1cm}
\vspace{0.0cm}
\caption{The \gg luminosity in the high-energy peak as a function of the electron beam energy for three laser wavelengths. Dashed curves: the same but \epem creation in the conversion region is switched off.}
\label{lumpeak} \vspace{-0.0cm}
\end{figure}

  \paragraph{The energy reach.} As mentioned in the introduction, the maximum value of the parameter $x$ should be below 4.8 in order to avoid \epem pair creation in the conversion region. The corresponding laser wavelength is $\lambda \approx 4.2 E_0 \;[\tev]\; \mum$. Reduction of the \gg luminosity due to pair creation is demonstrated in Fig.~\ref{lumpeak}, obtained by simulation; it shows the dependence  of the \gg luminosity in the high-energy peak on the electron beam energy for three laser wavelengths. One can see that a $1 \mum$ laser is good for $2E_0 < 550 \gev$, while a $2 \mum$ laser allows collider energies up to about $1100 \gev$ without any decrease in the luminosity due to pair creation. The ILC design energy is $2E_0=500 \gev$ with a further possible upgrade up to $1000 \gev$, so a laser with $2 \mum$ wavelength  will cover the whole ILC energy range. If the photon collider starts operation after the completion of \epem physics program, then the 2 \mum wavelength is the obvious choice.

\begin{figure}[!thb]
\centering
\hspace{-1.0cm} \includegraphics[width=10.0cm]{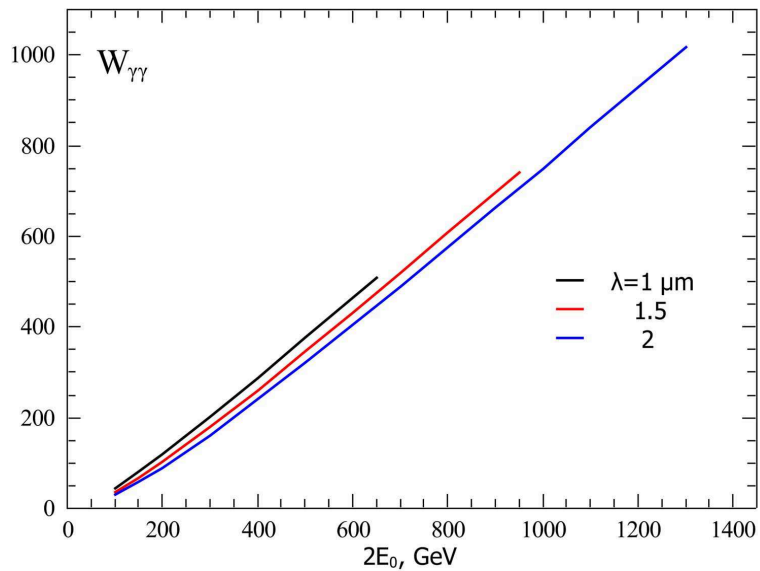}
\vspace{0.0cm}
\caption{The energy reach of the photon collider for three values of the laser wavelength. $W_{\gg}$ corresponds to the maximum of the \gg luminosity spectra (which is about 10\% lower than the edge energy given by Eq.~\ref{kin}).}
\label{en-reach} \vspace{-0.0cm}
\end{figure}

In the case of the ILC without an energy upgrade, the energy reach of the photon collider with $\lambda=2 \mum$ would be lower by about 13\% than with $\lambda=1 \mum$ (because the maximum value of $x$ would be a factor two lower, 2.38 instead of 4.75). The energy reach of the photon collider for the three wavelengths is shown in Fig.~\ref{en-reach}.
For the study of the Higgs boson and the top quark threshold with $\lambda=2 \mum$, one needs $2E_0=255$ and $550 \gev$, respectively, versus 210 and $485 \gev$ for $\lambda=1 \mum$; therefore, without an energy upgrade, the top threshold would be out of reach for the ILC.

\begin{figure}[!htb]
\centering
\hspace{-0.0cm} \includegraphics[width=9.0cm]{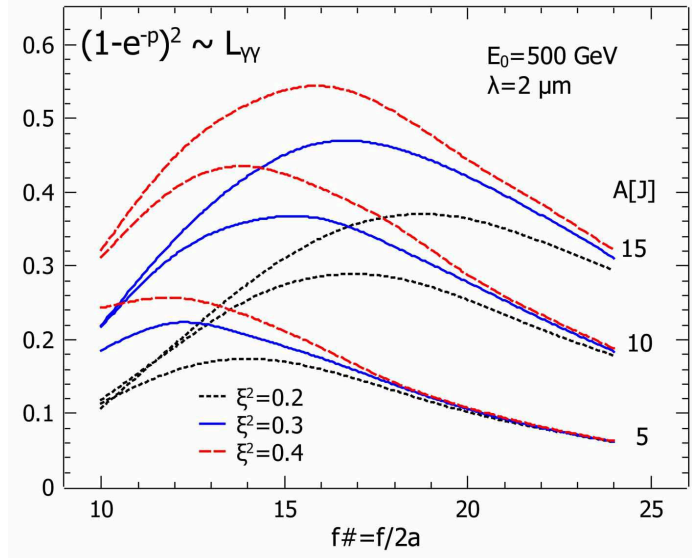}\hspace{-1cm}
\vspace{0.0cm}
\caption{Dependence of $L_{\gg}$ on the flash energy and $f_{\#}$
(flat-top laser beam) for several values of the parameter $\xi^2$.}
\label{conversion} \vspace{-0.cm}
\end{figure}

\paragraph{Laser flash energy.}  At present, the most powerful lasers have $\lambda \approx 1 \mum$. Lasers with $\lambda \sim 2 \mum$ do exist but are less powerful, in part because there has been no strong impetus for their development. Dependence of the \GG\ luminosity on the flash energy and
$f_{\#}=F/2R$ (flat-top laser beam) for several values of the parameter $\xi^2$ characterizing the multi-photon effects in Compton scattering (see Eq.~\ref{kin}) is shown in Fig.~\ref{conversion}. A flash energy of about 12 J is needed (for the optics geometry shown in Fig.~\ref{beams-quad}), which is about 20\% greater than with $\lambda=1 \mum$~\cite{PHOTON2005b} (nonlinear effects in Compton scattering are more important for longer wavelengths).

\paragraph{\gg luminosity spectra.}

\begin{figure}[!htb]
\centering
\hspace{-0.0cm} \includegraphics[width=7.cm]{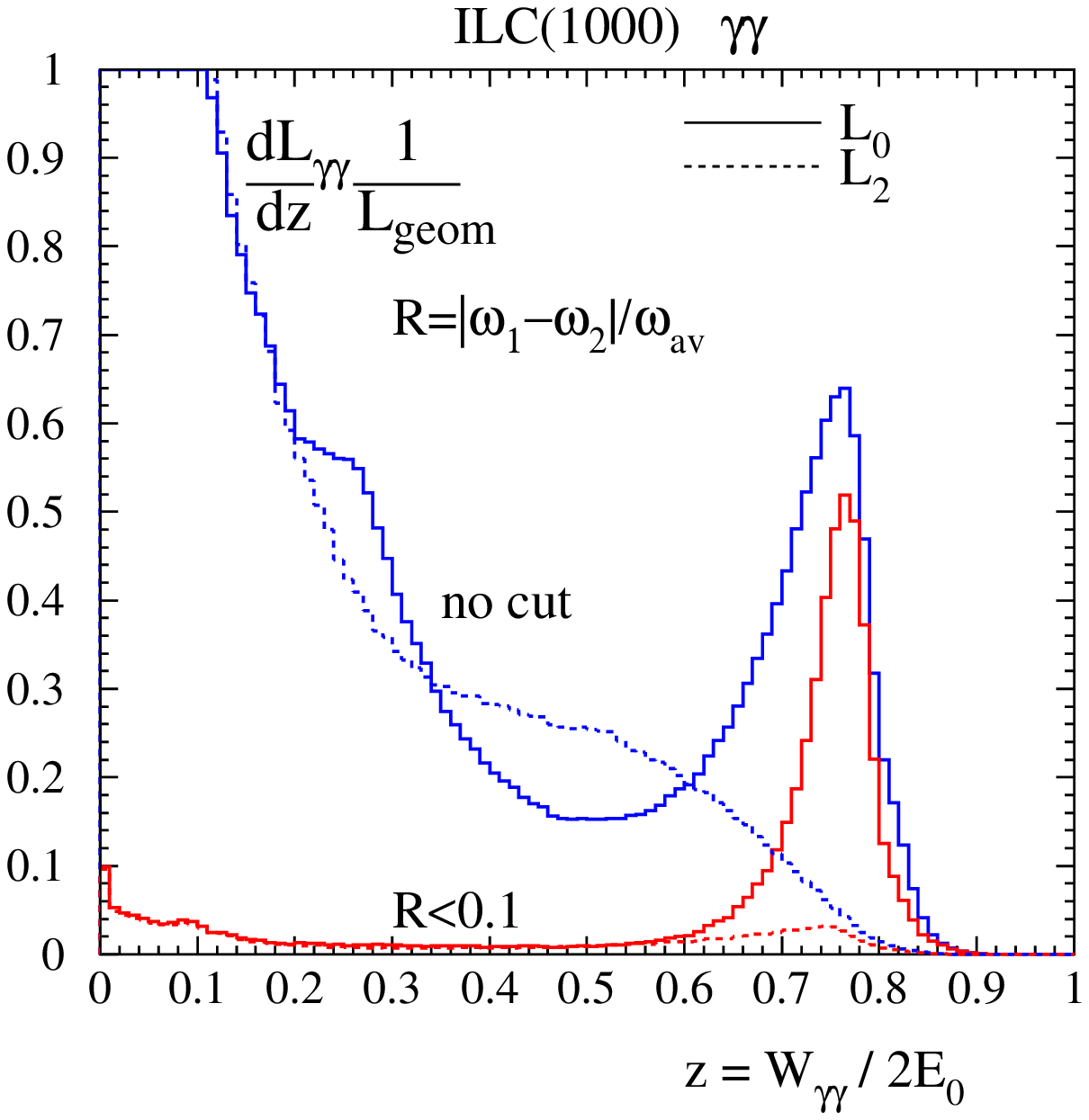}\hspace{-0cm} \includegraphics[width=7.cm]{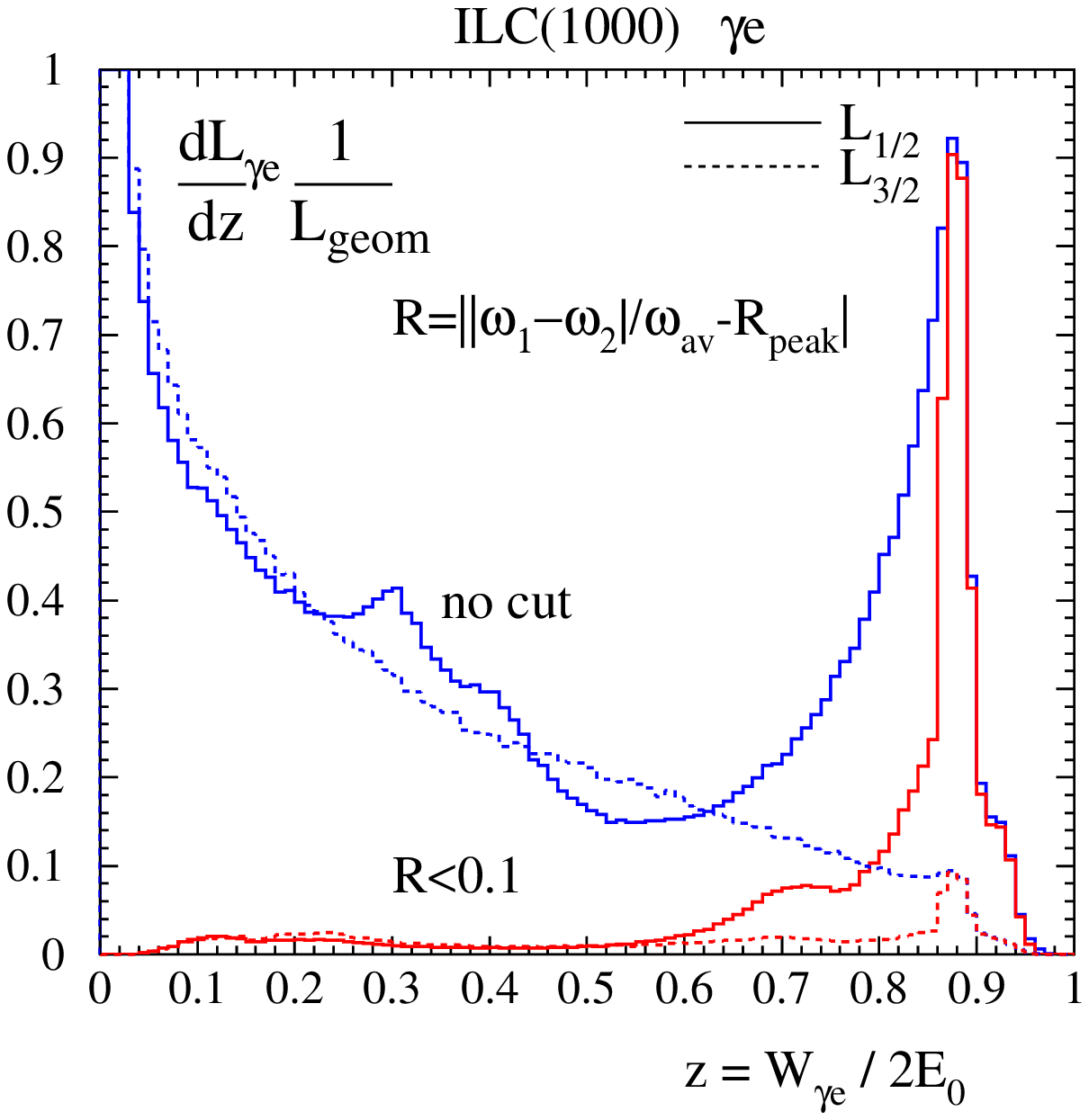}
\vspace{0.0cm}
\caption{The \GG\ (left) and \GE\ (right) luminosity spectra for typical ILC parameters at $2E_0=500 \gev$. Solid lines for $J_z$ of two colliding photons equal to 0, dotted lines for $J_z=2$ (1/2 and  3/2, respectively, in the case of \GE\ collisions). The total luminosity is the sum of the two spectra. Red curves (marked by $R<0.1$): with additional cut on the longitudinal momentum of the \gg system.}
\label{luminosity} \vspace{-0.cm}
\end{figure}

Luminosity spectra for \gg and \ge collisions at $2E_0=1 \tev$ with $\lambda=2 \mum$ are shown in Fig.~\ref{luminosity}. Such a \gg collider would be nice for the study of the (fake) diphoton resonance $\digamma(750)$ seen at the LHC in 2015.

\section{Discussion and conclusion}
  A unique feature of a high-energy \epem linear collider is the possibility to transform it into a photon (\gg and \ge) collider of comparable energy and luminosity. The incremental cost of a photon collider is negligible compared to the baseline LC cost; therefore, it has been considered for many years to be a natural part of any LC project. The recent observation at the LHC of a (fake) diphoton peak at $750 \gev$ reminded us once again that there are physics scenarios where the photon collider has big advantages; therefore, it would be wise to keep any LC design compatible with the photon collider option. In order to remove highly disrupted beams from the detector region, the photon collider needs a crossing angle somewhat larger than that for the \epem mode. In the ILC technical design, which is optimized for \epem, the crossing angle is $14 \mrad$, while the photon collider at ILC(500) needs a crossing angle of $25 \mrad$. Modification of the crossing angle following the completion of ILC's \epem program would be practically impossible from the technical standpoint; therefore, a consensus on the crossing angle is urgently needed.  In this paper, we analysed the factors that determine the optimal crossing angle for the photon collider and found that it can be decreased without degrading the \gg luminosity by increasing the laser wavelength from $1 \mum$ (optimum for $2E_0=500 \gev$) to $2 \mum$. In this case, a crossing angle of $20 \mrad$ is possible, which is acceptable for \epem as well.  The change of the laser wavelength from 1 to $2 \mum$ is advantageous for the photon collider at the ILC because such a laser would be a good fit for the entire energy range of the ILC, including an energy upgrade up to $2E_0 = 1 \tev$. 
  
  The increase of the crossing angle from 14 to 20 mrad will need 1.42 times larger bunch rotation by crab cavities. In the present ILC design the crab system consists of two 3.9 GHz 9-cell superconducting dipole cavities located 13.4 m from the IP~\cite{Adolphsen2007} which provides enough rotation for a 500 GeV beam. The total length of the cryomodule with two cavities is 2.3 m (while 3.8 m is reserved). The required RF power is low enough, about 3 kW. So, it seems there is no problem to add a third cavity to the cryomodule in order to adjust the crab system for 20 mrad crossing angle at $2E_0=1$ TeV. For $2E_0<700$ GeV no modification is needed. 

  In addition, the $20 \mrad$ crossing angle would be much more comfortable for a future high-energy \epem collider in the same tunnel, such as CLIC. At the CLIC, to avoid secondary background due to the coherent pairs created at the IP at highest energies (3 TeV), a crossing angle 20 mrad is required~\cite{Schulte2001}. This crossing angle also leads to acceptable multi-bunch effects due to parasitic collisions.
  
  This paper can be considered as a call for modification of the ILC crossing angle from 14 to $20 \mrad$.

\section*{Acknowledgements}

I would like to thank Kaoru Yokoya for his support of the photon collider and the help in finding a consensus on the crossing angle. 
The work was supported by the Ministry of Education and Science of the Russian Federation.

\end{document}